\documentclass[fleqn,10pt]{wlscirep}

\usepackage{amssymb}
\usepackage{amsmath}
\usepackage{amsthm}
\usepackage{url}  
\usepackage{graphicx}

\title{The prevalence of small world networks explained by modeling the competing dynamics of local signaling events in geometric networks}

\author[1,2,3]{Gabriel A. Silva}
\affil[1]{Department of Bioengineering, University of California San Diego, La Jolla California 92037} 
\affil[2]{Department of Neurosciences, University of California San Diego, La Jolla California 92037}
\affil[3]{Center for Engineered Natural Intelligence, University of California San Diego, La Jolla California 92037}

\affil[.]{Email: gsilva@ucsd.edu}

\begin{abstract}
Networks are ubiquitous throughout science and engineering. A number of methods, including some from our own group, have explored how one goes about computing or predicting the dynamics of networks given information about internal models of individual nodes and network connectivity, possibly with  additional information provided by statistical or descriptive metrics that characterize the network. But what can be inferred about network dynamics when there is no knowledge or information about the internal model or dynamics of participating nodes? Here, we explore how connected subsets of nodes competitively interact in order to activate a common downstream node they connect into. We achieve this by assuming a simple set of rules borrowed from neurophysiology. The model we develop reflects a local process from which global network dynamics emerges. We call this model a competitive refractory dynanics model. It is derived from a consideration of spatial and temporal summation in biological neurons, whereby summating post synaptic potentials (PSPs) along the dendritic tree contribute towards the membrane potential at the initial segment reaching a threshold potential. We first show how the 'winning node' or set of 'winning' nodes that achieve activation of a downstream node is computable by the model. We then derive a formal definition of optimized network signaling within our framework. We define a ratio between the signaling latencies on the edges of the network and the internal time it takes individual nodes to process incoming signals. We show that an optimal ratio is one where the speed of information propagation between connected nodes does not exceed the internal dynamic time scale of the nodes. We then show how we can use these results to arrive at a unique interpretation for the prevalence of the small world network topology in natural and engineered systems.
\end{abstract}
\begin{document}

\flushbottom
\maketitle
%
%
\thispagestyle{empty}

\section*{Introduction}
Networks are ubiquitous throughout science and engineering. A number of methods, including some from our own group, have explored how one goes about computing or predicting the dynamics of networks given information about internal models of individual nodes and network connectivity, possibly with  additional information provided by statistical or descriptive metrics that characterize the network \cite{buibas,Eld,Crev,kad,vogel}. But what can be inferred about network dynamics when there is no knowledge or information about the internal model or dynamics of participating nodes? In most situations, the number of hidden or unmeasurable variables (and time invariant parameters) in a model of a node far exceed the observable or measurable number. In the absence of accurate models of the participating nodes, inferring the dynamics of the network is not generally possible. Here, we explore how connected subsets of nodes competitively interact in order to activate a common downstream node they connect into. We achieve this by assuming a simple set of rules borrowed from neurophysiology. The model we develop reflects a local process from which global network dynamics emerges. We call this model a competitive refractory dynamics model. It is derived from a consideration of spatial and temporal summation in biological neurons, whereby summating post synaptic potentials (PSPs) along the dendritic tree contribute towards the membrane potential at the initial segment reaching a threshold potential. Upon activation and the generation of an action potential, subsequent PSP contributions, or external stimuli such as injected current, are not able to produce subsequent action potentials during the refractory period. 

We first show how the 'winning node' or set of 'winning' nodes that achieve activation of a downstream node is computable by the model. We then derive a formal definition of optimized network signaling within our framework. We define a ratio between the signaling latencies on the edges of the network and the internal time it takes individual nodes to process incoming signals. We show that an optimal ratio is one where the speed of information propagation between connected nodes does not exceed the internal dynamic time scale of the nodes. In other words, it represents a balance between how fast signals propagate through the network relative to the time needed for each node to process incoming information. An optimal ratio serves to maximize the amount of information individual edges between node pairs can support. A mismatch of this ratio leads to sub-optimal signaling and information flows in a network, and even a breakdown in signaling all together. We then show how we can use these results to arrive at a unique interpretation for the prevalence of the small world network topology in natural and engineered systems. While intuitively simple, the way we have formalized these concepts is allowing us to explore a number of applied questions in unique ways. 

The framework we develop assumes that three things can be observed or measured: First, the geometry, i..e. path integral, of the edges connecting the subset of nodes under consideration. Second, the signaling speed or conduction velocity of discrete signaling events propagating on the edges. Or equivalently, a signaling delay or latency of discrete signaling events on the edges. And third, a refractory period for activated nodes during which a recently activated node is rendered unable to respond to subsequent arriving signals.for some period of time. Given an observable starting state, the (possibly non-linear) combinatorial interactions associated with independent temporally offset signaling events propagating to differing degrees along their respective edges are responsible for the resultant dynamics.

\section*{Results}
We considered the geometrical construction of a network in the following sense. We assume that signals or discrete information events (e.g. action potentials in biological neurons) propagate between nodes along directed edges at a finite speed or conduction velocity, resulting in a temporal delay or latency of the signal arriving at the downstream node. Imposing the existence of signaling latencies implies a network that can be mapped to a geometric construction, where individual nodes could be assigned a spatial position in space in $\mathbb{R}^3$ for an ordered triplet $\bar{v}_i = \bar{x} = (x_1,x_2,x_3)$ for each vertex $v_i$ for all vertices $i = 1 \ldots N$. Where $N$ is the number of nodes and therefore the size of the network. Directed edges connecting node pairs could have a convoluted path integral, i.e. a Jordan arc. There is no restriction that edges have to be spatially minimizing straight line edges, i..e geodesics. A signaling latency $\tau_{ij}$ defines the ratio between the distance traveled on the edge relative to the speed of the propagating signal. For any set of connected vertex pairs $v_iv_j$ $\tau_{ij} = d_{ij} /s_{ij}$. While one does not have to explicitly consider $d_{ij}$ and $s_{ij}$, the existence of signaling latencies can always be mapped to these variables. This is analogous to the conduction velocity of action potentials traveling down the convoluted axon and axonal arborizations of a biological neuron. Note that formally, we consider graph models of such networks. In general we refer to the vertices and edges of graphs, but it should be understood that they model the nodes and connections of a networks. When there is no risk of confusion we use these terms interchangeably.  

The set of all geometric edges in the graph $G=(\bar{V},\bar{E})$ is given by $\bar{E}=\{e_{ij}\}$. The framework we develop however operates at a local scale. We define the subgraph $H_j$ as the tree graph that consists of all vertices $v_i$ with directed edges into $v_j$. We write $H_j(v_{i})$ to represent the set of all vertices $v_i$ in $H_j$ and $H_j[v_{i}]$ to refer to a specific $v_{i} \in H_j(v_{i})$. The edge set of $H_j$ is denoted by $H_j(\bar{E}_i)$ for the set of edges $e_{ij}$ into $v_j$. We assume there exist discrete signaling events traveling at a finite speed $s_{ij}$ on the edge $e_{ij}$. The signaling speed $s_{ij}$ from $v_i$ to $v_j$ is bounded such that $0 < s_{ij} < \infty$, i.e. it must be finite. By $H_j[v^*_{i}] \leadsto v_j$ we mean a vertex $v_{i} \in H_j$ that causally leads to the activation of $v_j$. We then define an absolute refractory period of a vertex $v_j$ by $R_j$. This reflects the internal dynamics of $v_j$ once a signaling event activates it. For example, the amount of time the internal dynamics of $v_j$ requires to make a decision about an output in response to being activated, or some reset period during which it cannot respond to subsequent arriving input signals. We place no restrictions on the internal processes or time that contribute toward $R_j$. We assume that we do not know and cannot observe the internal dynamic model of $v_j$, which could be quite complex with many hidden variables that produce its refractory period.  But we do assume we can observe it, in the sense that we can measure how long $R_j$ is. We also assume that $R_j > 0$, i.e. there cannot exist an infinitely fast or instantaneous recovery time, even though it can be arbitrarily short for any specific network. This is a reasonable assumption for any physically constructible network. 

\subsection*{Individual node states}
Consider a vertex $v_i$ with a directed edge to a vertex $v_j$. For $v_i$ to signal or communicate with $v_j$, there must be some physical signal representing a flow of information from $v_i$ to $v_j$ over the edge that connects them. This signal must travel at some finite speed $s_{ij}$. $s_{ij}$ could be a constant value for all edges, but this not need be true in the general case. Similarly, if all nodes $v_j$ in a network share the same internal dynamics, then $R_j = R \forall v_i \in \bar{V}$. But the framework does not assume this and can accommodate differing node specific values of the refractory period. In the construction of our framework, once $v_j$ receives a signal from $v_i$ it becomes refractory for a period $R_j$ and will not be able to respond to another incoming signal during this period of time. Because $R_j$ represents a period of time, note that as time progresses it shortens and eventually decays to zero, at which time $v_j$ is able to respond to another input.

We begin by letting $y_j(\Omega,t)$ represent the instantaneous state of vertex $j$ as a function of time and some (possibly unobservable) model with variable and parameter set $\Omega$. The internal state can be interpreted as a binary function at any time $t$ determined by $y_j(\Omega,t)$. We can define this function at some observation time $T_o$ as 

\begin{equation} \label{eq:yj}
 y_j(\Omega,T_o) =\begin{cases}
    1, & \text{iff $v_j$ can respond to an input}\\
    0, & \text{iff it is refractory to any input}
  \end{cases}
\end{equation} 

Once the winning node `activates' $v_j$ it will become refractory for a period of time $R_j$ during which $y_{j} =  0$, determined by its internal dynamic model. Importantly, note that if the state of $v_j$ at $T_o$ is $y_j = 0$ it could be refractory for some time $< R_j$ if had become refractory prior to $T_o$. This situation is interesting because we have to take into account phase shifts in $\tau_{ij}$ and $R_j$ at the sampling time $T_o$ in order to understand the dynamics of node activations. In other words, the timing of when arriving signaling events arrive at $v_j$ from nodes $v_i$ relative to the amount of the refractory period tine remaining for $v_j$ - the effective refractory period. In fact, this is at the core of the dynamical richness underlying distributed signaling in such geometric networks.

\subsection*{Signal flow dynamics for the sub-graph $H_j(v_i)$} \label{sec:deltaanalysis}
We begin by defining a simple relationship between $R_j$ and $\tau_{ij}$ that computes the state $y_j$ of $v_j$. The refraction ratio between the refractory period $R_j$ for vertex $v_j$ and the temporal signaling latency $\tau_{ij}$ into $v_j$ for a directed connected vertex $v_i$ on the edge $e_{ij}$ is given by 
\begin{equation}\label{eq:R}
\Delta_{ij} = \frac{R_j}{\tau_{ij}} = \frac{R_j \cdot s_{ij}}{d_{ij}}
\end{equation}
where $R_j > 0$.
Our analysis and enumeration of the combinatorial signaling space for $H_j(v_i)$ will proceed by a consideration of this ratio. 

There are a number of unallowable conditions that are necessitated by the physical construction of real world networks and the definitions above. $R =  0$ implies a non-refractory vertex capable of instantaneous recovery to an incoming signal from an upstream vertex, a condition which is not allowed. As $\tau_{ij} \rightarrow 0$ $\Delta_{ij}$ becomes undefined, which is equivalent to stating $d_{ij} \rightarrow 0$ since $\tau_{ij} \propto d_{ij}$ for a fixed signaling speed $s_{ij}$, with $s_{ij}$ of course as the constant of proportionality. The theoretical limit occurs when $R_j$ and $\tau_{ij} \rightarrow 0$. But this implies $s_{ij} = \infty$ or, functionally equivalently, $d_{ij} = 0$, i.e. no geometric distances to an edge, and simultaneously infinitely fast recovery times of the internal node refractoriness. But these conditions are unattainable. $\Delta_{ij}$ therefore necessarily implies finite dynamic signaling and information flow in a network, as required.

The trivial lower bound occurs as $R_j \rightarrow 0$, $y_i = 1$ $\forall$ $\tau_{ij}$. Intuitively, for any $v_i$ into $v_j$ when $y_i = 1$, the vertex with the shortest edge path integral will win and activate $v_j$. In other words, assuming a constant signaling speed $s_j \forall H_j(E)$ if we let $D_{ij} :=  \{d_{ij} : i = 1, 2, \dots N\}$ be the set of all edge path integrals for $H_j(E)$, then $H_j[v_{i}] \leadsto v_j = v_{i}(\min_i d_{ij})$ for $d_{ij} \in D_{ij}$. The trivial upper bound occurs as $R_j \rightarrow \infty$, $y_j = 0$ $\forall \tau_{ij}$, in which case there would be no information flow or signaling ever. 

\subsubsection*{Refraction ratio analysis with no temporal offset} \label{sec:notemp}
For any vertex pair $v_iv_j$ if $v_i$ where to signal exactly as $v_j$ becomes refractory then ideally $\tau_{ij}$ will be as small as possible but matched to $R_j$; in other words, it will be just a bit larger than $R_j$ so a signal from $v_i$ arrives at $v_j$ as soon as $v_j$ stops being refractory. The signal from the upstream node $v_i$ that reaches $v_j$ first will `win' and activate $v_j$ and make it refractory to other arriving signals for a period $R_j$. We formalize these ideas in the following way. Begin by considering any two vertices $v_i$ and $v_j$ in a complete directed geometric graph $G(\bar{V},\bar{E})$. Consider what happens when $v_i$ signals $v_j$ at time $t = t_i$. The shortest physically possible reaction time for $v_j$ in all cases will be a signal reaching it from $v_i$ just as its refractory period is ending. This occurs when $\tau_{ij} \rightarrow R_j^+$, i.e. approaches $R_j$ from the right, that is, is slightly longer than $R_j$. Let $\Delta^o_{ij}$ represent the set of all $\Delta_{ij}$ ratios for all $v_i$ vertices with directed edges $e_{ij}$ into $v_j$ that belong to $H_j(v_i)$ for which the condition $\tau_{ij} \rightarrow R_j^+$ is met: 
\begin{equation} \label{eq:col}
\Delta^o_{ij} := \{\Delta_{ij} : i = 1, 2, \dots N | \Delta_{ij} \text{ for } \tau_{ij} \rightarrow R_j^+\}
\end{equation}
This then implies that 
\begin{equation*}
\forall \Delta_{ij} \in \Delta^o_{ij} \implies \Delta_{ij} < 1
\end{equation*}

We can then prove the following relation: Let $t_i \forall i \in \Delta^o_{ij} = t_o$, i.e. all vertices into $v_j$ initiate a signaling event at the same time $t_o$. Assume $v_j$ becomes refractory exactly at $t_o$. For any refractory period $R_j$ the winning refraction ratio $\Delta^*_{ij}$ for the 'winning' vertex $H_j[v^*_i] \leadsto v_j$ is given by 
\begin{equation} 
\label{eq:rwithnophi}
\Delta^*_{ij} = \max (\Delta_{ij}) \text{ for } \Delta_{ij} > 0 \in \Delta^o_{ij}
\end{equation} 

\begin{proof}
Given the set $\Delta^o_{ij}$, assume it is a well ordered set, i.e. there exists a smallest element in $\Delta^o_{ij}$. Then order the elements $(\Delta_{ij}) \in \Delta^o_{ij}$ from smallest to largest for an index $k = 1, 2, \dots k$. In the limit as $\tau_{ij} \rightarrow R^+_j$ the winning vertex $v^*_i$ will be the vertex with the refraction ratio $\Delta^*_{ij} = (\Delta_{ij})_k = \max (\Delta_{ij})$ as required. For any $R_j$ value the numerator for all $\Delta_{ij}$ can be reduced to unity (i.e. factored out) because it is common to all $\Delta_{ij}$. Then, the largest ratio will be the ratio with the smallest denominator, which represents the shortest latency $\tau_{ij}$ and therefore the winning vertex $v_i$. Because of how the set of $ \Delta_{ij}$ are ordered, from smallest to largest, it ensures that the condition in equation \ref{eq:rwithnophi} results in the winning vertex.
\end{proof}
In addition, we make the following observation about the limit the value $\Delta_{ij}$ can take. Given the set $\Delta^o_{ij}$, $\forall \Delta_{ij} \in \Delta^o_{ij}$, if we let $\Delta_{ij,max}$ represent the largest attainable value by an element of $\Delta^o_{ij}$, 
\begin{equation} 
\label{eq:Rapp}
\Delta_{ij,max} = \lim_{\tau_{ij} \rightarrow R^+_j} \frac{R_j}{\tau_{ij}} \rightarrow 1
\end{equation}
This follows directly from the definition of $\Delta_{ij}$ in equation \ref{eq:R}. 

Alternatively, we can re-write equation \ref{eq:rwithnophi} as an inequality condition of a difference.
Let $t_i \forall i \in \Delta^o_{ij} = t_o$. Assume $v_j$ becomes refractory exactly at $t_o$. The 'winning' vertex $H_j[v^*_i] \leadsto v_j$ is given by
\begin{equation} 
\label{eq:rwithnophi2}
H_j[v^*_{ij}] = v_{i} \in H_j \text{ such that } \min \lbrack (\tau_{ij} - R_j) > 0 \rbrack 
\end{equation}

\begin{proof}
The necessary condition for $H_j[v^*_i] \leadsto v_j$ is $\tau_{ij} \rightarrow R_j^+$. In the limit $\Delta_{ij} \rightarrow \Delta_{ij,max}$ when $\tau_{ij} \rightarrow R_j$. This implies that $(\tau_{ij} - R_j) \rightarrow 0$. Therefore, in every case $v^*_{i}$ will be the smallest positive value of $(\tau_{ij} - R_j) \rightarrow 0$, or $\min_i \lbrack (\tau_{ij} - R_j) > 0$ as required. Note that the condition for positive values of the difference is necessary because negative values imply that the signal from $v_{i}$ arrive at $v_j$ while it is still refractory.
\end{proof}
Algorithmically, equation \ref{eq:rwithnophi2} is much more efficient to implement because one only needs to compute a difference compared to equation \ref{eq:rwithnophi} which necessitates computing a ratio. This becomes significant when computing in parallel all $H_j \in G(\bar{V},\bar{E})$. 

\subsubsection*{Refraction ratio analysis with temporal offset}
Under most conditions, there is likely to be a temporal offset between when each $v_i$ signals at $t_i$ and how far along $v_j$ is in its recovery from its refractory period due to a previous signaling event relative to an observation time $T_o$. This would be the case for all situations other then when $v_j$ becomes refractory exactly at $t_i$. We first consider the case where all $v_i$ signal $v_j$ at the same $t_i$ for every $v_i \in H_j$, that is, when $t_i \forall i \in \Delta^o_{ij} = t_o$. 
Let $\phi_j$ represent a temporal offset from $R_j$, such that at $t_i$ 
\begin{equation} \label{eq:rbar}
\bar{R}_{j} = R_j - \phi_j \text{ where } 0 \leq \phi_j \leq R_j
\end{equation} \label{eq:effectiveR}
We call $\bar{R}$ the effective refractory period. It reflects the amount of time remaining in the recovery from $R_j$ at the time $t_o$. We re-write equation \ref{eq:R} as
\begin{equation} \label{eq:Rbar}
\bar{\Delta}_{ij} = \frac{\bar{R}_j}{\tau_{ij}} = \frac{\bar{R}_j \cdot s}{d_{ij}}
\end{equation}

When $\phi_j = 0$ it implies $v_j$ became refractory exactly when $v_i$ signaled at $t_i$, which here we are assuming all $v_i$ signal at the same time $t_o$. This is effectively the special case described by equations \ref{eq:rwithnophi} and  \ref{eq:rwithnophi2}. When $\phi_j = R_j$ it implies that $v_j$ is not refractory and can respond to an input from any $v_i$ at any time. Note how when $\phi_j = R_j$ $v_j$ may have been refractory at some time $t \leq t_o -R_j$, but assures the condition that $\bar{R}_j = 0$ at $t_o$. 

Furthermore, the following then applies.
Let $t_i \forall i \in \Delta^o_{ij} = t_o$. If $\phi_j = R_j$ then the edge path integral for the 'winning' vertex $H_j[v^*_i] \leadsto v_j$ with $v^*_i(\Delta^*_{ij})$ will be $\min(d_{ij}) \forall d_{ij} \in \Delta^o_{ij}$.
\begin{proof} 
Given the definition of $\Delta_{ij}$ in equation \ref{eq:R}, for a constant $s_{ij}$, $\Delta^*_{ij} = \lim_{\tau_{ij} \rightarrow R^+_j} \max(\Delta_{ij})$ when $\tau_{ij} \in \Delta^o_{ij} = \min(\tau_{ij})$ for $v^*_i$, since if $\phi_j = 0 \forall i \in H_j(v_i)$ $y_j = 1 \forall i \in H_j[v_{ij}]$. This condition will be met when $d_{ij} \in \Delta^o_{ij} = \min(d_{ij})$.
\end{proof}
And by direct extension of of the results above, we can write the equivalent expressions for $0 \leq \phi_j \leq R_j$.

Let $t_i \forall i \in \bar{\Delta}^o_{ij} = t_o$. Assume $v_j$ has an effective refractory period given by $\bar{R}_j$ for some value of $\phi_j$ at $t_o$. The refraction ratio $\bar{\Delta}^*_{ij}$ for the 'winning' vertex $H_j[v^*_i] \leadsto v_j$ with $v^*_i(\bar{\Delta}^*_{ij})$ is given by 
\begin{equation} 
\label{eq:rwithphi}
\bar{\Delta}^*_{ij} = \max(\bar{\Delta}_{ij}) \text{ for } \bar{\Delta}_{ij} > 0 \in \bar{\Delta}^o_{ij}
\end{equation} 

Given the set $\bar{\Delta}^o_{ij}$, $\forall \bar{\Delta}_{ij} \in \bar{\Delta}^o_{ij}$, and assuming $v_j$ has an effective refractory period given by $\bar{R}_j$ for some value of $\phi_j$ at $t_o$, if we let $\bar{\Delta}_{ij,max}$ represent the largest attainable value by an element of $\bar{\Delta}^o_{ij}$, 
\begin{equation} 
\label{eq:Rapp2}
\bar{\Delta}_{ij,max} = \lim_{\tau_{ij} \rightarrow \bar{R}^+_j} \frac{\bar{R}_j}{\tau_{ij}} \rightarrow 1
\end{equation}

And finally, assume $v_j$ has an effective refractory period given by $\bar{R}_j$ for some value of $\phi_j$ at $t_o$. The 'winning' vertex $H_j[v^*_i] \leadsto v_j$ is given by
\begin{equation} 
\label{eq:rwithnophi2b}
H_j[v^*_{ij}] = v_{i} \in H_j \text{ such that } \min \lbrack (\tau_{ij} - \bar{R}_j) > 0 \rbrack
\end{equation}

In the most general case $t_i$ for $v_i \in H_j$, i.e. the times at which each vertex initiates a signal, would not be expected to be all the same. One would expect that $t_i \neq t_o$ $\forall i$. At any given arbitrary observation time $T_o$ a signal from any $v_i$ may be traveling part way along $e_{ij}$ at a speed $s_{ij}$, effectively shortening $\tau_{ij}$. Or it may be delayed in signaling if $v_i$ signals some time after $T_o$, effectively lengthening $\tau_{ij}$. At time $T_o$ we need to take into account the degree of signaling progression for each edge and define $\bar{\tau_{ij}}$ analogous to $\bar{R}$. Note that we write $T_o$ to distinguish the case when $t_i \forall i \in \bar{\Delta}^o_{ij} \neq t_o$ since $t_i$ here represents the time that vertex $v_i$ signals, which could be different than the time $T_o$ at which the network is observed. 

We then extend the latency definition as follows. Let $\tau_{ij}$ represent the temporal delay, the latency period, for a signal that travels on the edge $e_{ij}$ for vertex $v_i \in H_j$ when $v_i$ initiates a signaling event at $t_i$ which could come before, right at, or after the instantaneous observation time $T_o$, $ t_i \geq T_o$ or $ t_i < T_o$. We then define a temporal offset for $\tau_{ij}$, an effective shortening or lengthening of $\tau_{ij}$ as follows 
\begin{equation} \label{eq:taubar}
\bar{\tau}_{ij} = \tau_{ij} + \delta_{ij} \text{ where, }  \delta_{ij} \in \mathbb{R} 
\end{equation}
We extend equation \ref{eq:Rbar} as 
\begin{equation} \label{eq:Rbar2}
\Lambda_{ij} = \frac{\bar{R}_j}{\bar{\tau}_{ij}}
\end{equation}

$\delta_{ij} > 0$ represents an effective delay or elongation beyond $\tau_{ij}$. In other words, it represents the vertex $v_i$ initiating a signal at some time after $T_o$. Values $-\tau_{ij} < \delta_{ij} < 0$ represent an effective shortening of $\tau_{ij}$. This would be the case when $v_i$ had initiated a signal that was traveling part way along the edge $e_{ij}$ towards $v_j$ prior to the sampling time $T_o$. When $\delta_{ij} = 0$ it implies that $v_i$ signals exactly at the moment the network is observed. And when $\delta_{ij} = -\tau_{ij}$ it implies that the signal arrives at $v_j$ at the moment the network is observed. Values of $\delta_{ij} < \tau_{ij}$, which result in $\bar{\tau}_{ij} < 0$, represent a signal arriving at $v_j$ \emph{prior} to the observation time $T_o$. 

For completeness, we re-write equation \ref{eq:yj} to include $\bar{R}_j$ and $\bar{\tau}_{ij}$ as  
\begin{equation*} \label{eq:yj2}
 \bar{y}_j(\Omega,T_o) =\begin{cases}
    1, & \text{iff $v_j$ can respond to an input from any $v_i$}\\
    0, & \text{iff it is refractory to any input for a period $\bar{R}_j$ that begins at $T_o$}
  \end{cases}
\end{equation*} 

We also similarly define 
\begin{equation} \label{eq:BigLSet}
\Lambda^o_{ij} := \{\Lambda_{ij} : i = 1, 2, \dots N | \Lambda_{ij} \text{ for } \bar{\tau}_{ij} \rightarrow \bar{R}_j^+\}
\end{equation}
as a well ordered (smallest to largest) set of refraction ratios analogous to equation \ref{eq:col}.

We can now formally state the conditional relationship between $\bar{\tau}_{ij}$ and $\bar{R}_j$ necessary for signaling in the general case.

Let $G=(\bar{V},\bar{E})$ represent a complete geometric graph model of a network consisting of subgraphs $H_j(v_i)$ with directed edges $H_j(\bar{E})$ into vertex $v_j$. Assume a signaling speed $s_{ij}$ between $v_i$ and $v_j$. $v_i$ may activate $v_j$ iff $\bar{\tau}_{ij} > \bar{R}_j$.

\begin{proof}
$v_j$ will be in a state where it is capable of being activated in response to receiving a signal from $v_i  \in H_j(v_i)$ only when $\bar{y}_j(\Omega,T_o) =1$. This is the case only following the completion of the effective refractory period $\bar{R}_j$. At an observation time $T_o$ $\bar{R}_j$ can take on any value over its range of $0 \leq \bar{R}_j \leq R_j$. In order for the signal from $v_i$ to arrive at $v_j$ when $\bar{y}_j(\Omega,T_o) =1$ then, in every case $\bar{\tau}_{ij} > \bar{R}_j$.
\end{proof}

And by direct extension of the results above we can write 

If $\phi_j = R_j$ then $\bar{R}_j = 0 \Rightarrow \Lambda_{ij} =0$ and the 'winning' vertex $H_j[v^*_i] \leadsto v_j$ will be given by the vertex with the delay $\min(\bar{\tau}_{ij}) \forall d_{ij} \in \Lambda^o_{ij}$. 

We then arrive at the general theorems that completely describe the competitive refractory dynamics framework for each subgraph $H_j(v_i)$ that makes up the network. \newline

\textbf{Theorem 1.} Assume $t_i \forall i \in \Lambda^o_{ij}$ at $T_o$ for some value of $\delta_{ij}$, such that all $t_i$ need not necessarily be equivalent. Assume $v_j$ has an effective refractory period given by $\bar{R}_j$ for some value of $\phi_j$ at $T_o$.  The refraction ratio $\Lambda_{ij}$ for the 'winning' vertex $H_j[v^*_i] \leadsto v_j$ is given by 
\begin{equation} 
\label{eq:rwithphi}
\Lambda^*_{ij} = \max(\Lambda_{ij}) \text{ for } \Lambda_{ij} > 0 \in \Lambda^o_{ij}
\end{equation} 

\begin{proof}
The proof parallels the proofs outlined above. Given the set $\Lambda^o_{ij}$, assume it is a well ordered set. Then order the elements $(\Lambda_{ij}) \in \Lambda^o_{ij}$ for an index $k = 1, 2, \dots K$. In the limit as $\bar{\tau}_{ij} \rightarrow \bar{R}^+_j$ the winning vertex $v^*_i(\Lambda^*_{ij})$ will be the vertex associated with the refraction ratio $(\Lambda_{ij})_k = \max(\Delta_{ij})$ as required. In other words, in every case $H_j[v^*_i] \leadsto v_j$ will always be the vertex that arrives at $v_j$ first subject to it recovering from its effective refractory period $\bar{R}_j$ given the temporal evolution of $\bar{\tau}_{ij}$ for all $i$ and $\bar{R}_{j}$ at the instantaneous time $T_o$ at which the network is observed. 
\end{proof}

\textbf{Theorem 2.} Given the set $\Lambda^o_{ij}$, $\forall \Lambda_{ij} \in \Lambda^o_{ij}$, with $v_j$ having an effective refractory period $\bar{R}_j$ for some value of $\phi_j \neq R_j$ at $T_o$ such that $\bar{R}_j \neq 0$, if we let $\Lambda_{ij,max}$ represent the largest attainable value by an element of $\Lambda^o_{ij}$, 
\begin{equation} 
\label{eq:Rapp2}
\Lambda_{ij,max} = \lim_{\bar{\tau}_{ij} \rightarrow \bar{R}^+_j} \frac{\bar{R}_j}{\bar{\tau}_{ij}} \rightarrow 1
\end{equation}

\begin{proof}
The proof follows directly the definition and proofs of equations \ref{eq:Rbar} and \ref{eq:taubar}.
\end{proof}

If $\Lambda_{ij} = 0$ then we immediately know that $v_{i}$ with $\min({\bar{\tau}_{ij}) \forall v_i \in H_j(v_i})$ will win since $\Lambda_{ij} = 0$ only when $\bar{R}_j = 0$ which implies that $\bar{y}_j(\Omega,t) = 1$. \newline

\textbf{Theorem 3.} For each $v_i \in H_j(v_{ij})$ assume a $\bar{\tau}_{ij}$ at an observation time $T_o$. If the effective refractory period for $v_j$ at $T_o$ is $\bar{R}_j$, then the 'winning' vertex $H_j[v^*_i] \leadsto v_j$ is given by
\begin{equation} 
\label{eq:rwithnophi2b}
H_j[v^*_{ij}] = v_{i} \in H_j(v_{ij}) \text{ such that } \min \lbrack (\bar{\tau}_{ij} - \bar{R}_j) > 0 \rbrack
\end{equation}

\begin{proof}
The necessary condition for $H_j[v^*_i] \leadsto v_j$ is $\bar{\tau}_{ij} \rightarrow \bar{R}_j^+$. By theorems 1 and 2, in the limit $\Lambda_{ij} \rightarrow \Lambda_{ij,max}$ when $\bar{\tau}_{ij} \rightarrow \bar{R}_j$, which represents the largest value attainable by $v^*_{ij}$. This implies that $(\bar{\tau}_{ij} - \bar{R}_j) \rightarrow 0$. Therefore, in every case $v^*_{ij}$ will be the smallest positive value of  $(\bar{\tau}_{ij} - \bar{R}_j) \rightarrow 0$, or $\min\lbrack (\bar{\tau}_{ij} - \bar{R}_j)) > 0$ as required. In this case also, note that the condition for positive values of the difference is necessary because negative values imply that the signal from $v_{ij}$ arrive at $v_j$ while it is still refractory.
\end{proof}

\subsubsection*{Inhibitory inputs into $v_j$}
We can extend the framework to include inhibitory inputs from $v_i$ in the following intuitive way: Given a 'winning' vertex $v_i$ $H_j[v^*_i] \leadsto v_j$, $v_j$ generates an output signal in turn if the input from $v_i$ was excitatory or does not produce an output but becomes refractory nonetheless if the input from $v_i$ was inhibitory.

\subsection*{Explicit contribution of the internal processing time of $v_i$ to $\bar{\tau}_{ij}$} \label{sec:tjmag}
It is possible that upon the activation of a node there exists a finite period of processing time associated with its internal dynamics required to make a decision prior to the output of a signal. This is a function of the internal dynamic model of the node independent of the network dynamics. The details of such an internal model can be node specific or specific to the physical system being modeled as a network. But the practical consequence on the network dynamics as we develop them here is a contribution to the effective latency of the signaling event reaching downstream nodes. This is strictly different and independent from, i.e. can evolve in parallel with, the effective refractory period of that begins at the moment the node generating the output signal is activated. If this internal processing time approaches the time scale of the signaling speed and refractory periods then it will affect the dynamics of the network. If however, it is much smaller than both then it could have a negligible effect and be ignored. In the limit as it approaches zero it reflects a (near) instantaneous turn around time between when a signal that activates the node arrives and when that node outputs a signal in turn. The conditions that properly model any specific network are subject to the physical details of the system. However, the framework can take this into account explicitly.  

Before continuing we note that there is a potential source of confusion with regards to the node subscript notation that the reader has to be aware of. This internal processing time is a property of node $v_j$ as a 'winning' signal activates it as per the competitive refractory model. However, it affects the signaling latency of a signal outputted by $v_j$ traveling along the edges it connects to downstream nodes. But in the context of signals leaving $v_j$ its subscript in effect changes to $v_i$ because it is now an input into the nodes it connects to. The subscript notation is of course relative to the context of the individual $v_i$,$v_j$ node pairs under consideration. This is the mathematical equivalence of the relative usage of the terms 'presynaptic' and 'postsynaptic' when considering biological neuronal signaling. We attempt to be as clear and explicit as possible regarding this distinction in the mathematical description that follows. 

Let $t_j$ be the time at which an activated node $v_j$ outputs a signal. With a slight abuse of notation we write $|t_j|$ to define the magnitude or period of the internal dynamic processing time by $v_j$. This is the time between when a 'winning' signal $H_j[v^*_i] \leadsto v_j$ arrives at $v_j$ with a latency $\bar{\tau}_{ij}$ starting at an observation time $T_o$, or equivalently a latency $\tau_{ij}$ if $T_o = t_i$, and when $v_j$ actually sends out an output. 

For simplicity, up to this point we have implicitly assumed that $|t_j| = 0$, reflecting an instantaneous output at the moment that $v_j$ is activated. $R_j$ must begin anywhere between the moment of activation of $v_j$ when $\bar{\tau}_{ij}$ for the winning node ends and $t_j$. But this is strictly a property of the specifics of the physical system being modeled as a function of the internal dynamics of $v_j$. If $|t_j| > 0$ then it will affect the time at which it initiates an out going signal. Recall that for any vertex $v_i$ we denote the time at which it initiates an out going signal as $t_i$. This is where the switch in the index notation occurs. For an activated vertex $v_j$ with $|t_j| > 0$ we write here its signal initiating time as $t'_i$ simply to indicate that this particular $t_i$ corresponds to the activated node $v_j$.

From a computational perspective we can absorb the effect of $|t_j| > 0$ by an appropriate elongation in $\delta_{ij}$. We formalize this notion in the following way. We can re-write \ref{eq:taubar} such that for any vertex with $|t_j| > 0$ we express the effective latency $\bar{\tau}_{ij}$ as
\begin{equation}
\bar{\tau}_{ij} = \tau_{ij} + \delta'_{ij} + |t_j| = \tau_{ij} + \delta_{ij}
\end{equation} 

\begin{proof}
$\tau_{ij}$ is a computed quantity dependent on $s_j$ and $d_{ij}$. But computing $\bar{\tau}_{ij}$ involves taking into account $\delta_{ij}$ such that $\bar{\tau}_{ij} = \tau_{ij} + \delta_{ij}$ \emph{c.f.} equation \ref{eq:taubar}. $\delta_{ij}$ affects when $v_j$ outputs or initiates a signal. The effect of $|t_j| > 0$ is to delay by some constant value when that occurs at time $t'_i$.   
\end{proof}

Because $\delta_{ij} \in \mathbb{R}$ it absorbs $\delta'_{ij}$, which we use to write the component of $\delta_{ij}$ that does not account for the extra time due to $|t_j|$. There are three possible conditions in which this can be the case. 1. a signal $H_j[v^*_i] \leadsto v_j$ arrives at $v_j$ after which some amount of internal processing time given by $|t_j|$ $v_j$ makes a decision to output a signal in turn and does so instantaneously following that decision. 2. a signal $H_j[v^*_i] \leadsto v_j$ arrives at $v_j$ which makes an instantaneous decision to output a signal in turn and it takes $|t_j|$ before the signal from $v_j$ actually goes out. Or 3. a combination of scenarios 1. and 2. whereby some fraction $|t_j|/A$ of $|t_j|$ represents the time required to make a decision and $|t_j|/B = |t_j| - (|t_j|/A)$ represents the time between when a decision is made and an output signal actually goes out. Note of course that independent of the magnitude of $|t_j|$ the ratio $A/[B(A-1)] = 1$. From a practical perspective, we do not need to distinguish between the three conditions, which will be system (network) specific. What matters to us is that if $v_j$ produces a signal due to $H_j[v^*_i] \leadsto v_j$ then $|t_j|$ will effectively cause an elongation of $\bar{\tau}_{ij}$. If however, for a specific system $|t_j| << \tau_{ij} \text{ and } R_j$ it would have no effect on the dynamics. For any real physical system $|t_j|$ must always be finite and greater than zero of course, but we can safely ignore its effects if its time scale is much shorter than then the dynamics of the network. 

The computation of the refraction ratio given by equation \ref{eq:Rbar} above
\begin{equation*} \label{eq:Rbar2}
\Lambda_{ij} = \frac{\bar{R}_j}{\bar{\tau}_{ij}}
\end{equation*}
and its associated theorems do not change. What does change is that in computing $\Lambda_{ij}$ the internal dynamics of $v_j$ given $|t_j|$ through its effect on elongating $\bar{\tau_{ij}}$ will affect the value of $\Lambda_{ij}$ and therefore the network dynamics, if $|t_j|$ is on the scale of $\tau_{ij} \text{ and } R_j$ .

\subsection*{Probabalistic extension of the framework}
The description of the framework we outline above is deterministic in that one individual $v_i$ node is capable of activating $v_j$, $H_j[v^*_i] \leadsto v_j$. Given an activation by $v_i$ i$v_j$ is guaranteed to produce an output if the input from $v_i$ was excitatory or does not produce an output but becomes refractory if the input from $v_i$ was inhibitory. This is equivalent to stating that the probability of $v_j$ responding to a 'winning' signal from a $v_i$ is unity. However, we can extend these concepts to add a probabilistic component to the framework.

To achieve this, we assign a probability distribution $P_i$ to the likelihood of activation of $v_j$ for a winning vertex $v_i$: $H_j[v^*_i] \leadsto v_j$, with $v_j(P_i)$ indicating the output probability of $v_j$ for some output probability threshold $P_{threshold}$ given that the condition $H_j[v^*_i] \leadsto v_j$ has occurred and a specific value of $P_i$ for the that particular instance of a winning event $[v^*_i]$:
\begin{equation} \label{eq:vj1}
 v_j(P_i) =\begin{cases}
    1, & \text{iff } H_j[v^*_i] \leadsto v_j \text{ and } v^*_i(P_i) \geq P_{threshold}\\
    0, & \text{if } H_j[v^*_i] \leadsto v_j \text{ and } v^*_i(P_i) < P_{threshold}
  \end{cases}
\end{equation} 
It is critical to understand that equation \ref{eq:vj1} does not specify \emph{what} the output from $v_j$ will be given $v_j(P_i) =1$, which could be an actual signal if $v^*_i$ is excitatory or no signal if it is inhibitory, but only that $v_j$ \emph{does} respond in some way to $v^*_i$ at $t_i + \tau_{ij}$, i.e. it either outputs a signal and becomes refractory or does not output a signal and becomes refractory. Also note that when $v_j(P_i) =0$ we state the condition as a conditional 'if' statement ($if$ and not $iff$) because there exist other conditions outside those expressed by equation \ref{eq:vj1} that can result in no output from $v_j$ being possible, i.e. $v_j(P_i) =0$, namely, when $v_j$ is refractory. 

\subsection*{Summation from fractional contributions of multiple winning nodes: geometric dynamic percptrons}
We can further extend the framework so that instead of a single node $v_i$ activating $v_j$, whether deterministically or probabilistically, we now consider a 'running' summation of contributions from a number of $v_i \in H_j(v_i)$ adding up to a threshold value that then activates $v_j$. Upon activation $v_j$ becomes refractory as previously described. Conceptually this represents a competitive refractory geometric dynamic extension of the classical notion of a perceptron. It naturally extends the notion of the classical perceptron in several ways. Conceptually, our perceptrons display a geometric morphology to account for the computed latencies on the edges that represent the inputs into $v_j$. The critical interplay between the latencies and timing of the signaling dynamics on the input edges and the evolving refractory state of $v_j$ explicitly determine the running summation towards threshold of signal contributions from arriving inputs. We define a decay function (below) that provides a memory or history for previous arriving signals to affect the running summation, thereby resulting in diminishing but non-zero contributions from inputs at $v_j$ at any given instantaneous moment for offset arriving signals. The computational prediction being made here is not which $v_i \in H_j(v_i)$ will activate $v_j$ at what time $T_o + \tau_{ij}$, but what \emph{subset} of $H_j(v_i)$ will do so at some $T_o + \Delta t$. Beyond the immediate scope of the description we introduce here, this perceptron model produces a very rich dynamics and significantly increased number of states relative to traditional models. 

Consider an observation time $T_o$, and assume that $v_j$ is not refractory. As a function of  time the 'running' summation $\Sigma_r$ from $H_j(v_i)$ must reach a threshold $\Sigma_T$ in order for $v_j$ to activate at some time $t \geq T_o$. Once activated, $v_j$ becomes refractory for a period $R_j$ as usual. The specific contribution from one $v_i \in H_j(v_i)$ will be the value of the weight (synaptic strength) corresponding to that node, $w_{ij}$. As is typical in a perceptron, there is a set of weights associated with all incoming connections $W_j = \{w_{ij}\}$. The maximum value of the contributing weight $w_{ij,max}$ occurs at the time that the signal from $v_i$ arrives at $v_j$ after the end of the relative refractory period for $v_j$, $\bar{R}_j$. This occurs at the time $(\bar{\tau}_{ij} - \bar{R}_j)$. After which, i.e. beginning at the next time step, the contributing value of the weight begins to decay as a function of time: $w_{ij} < w_{ij,max}$ for times $t > (\bar{\tau}_{ij} - \bar{R}_j)$. In other words, there is a finite memory at future times to the arrival of a given signal, scaled to its weight value, that progressively decays over time to zero. This produces a complex and dynamic interplay between how far along the respective signals are along their edges towards $v_j$ relative to each other at any given moment in time (signaling latencies- which encode an underlying geometry to the network), the magnitude of the contribution from the respective weights once they do arrive, the kinetics of the decay of the contributing weights as a function of time, and the timing of the refractory state and recovery from refractory state of $v_j$ that all contribute towards $\Sigma_T$. And as above, we also account for excitatory and inhibitory weights in the sense that inhibitory weights subtract from $\Sigma_r$. This dynamics models the canonical notion of temporal summation of post synaptic potentials at the dendritic tree of a neuron. This model extends the conceptual construction of a perceptron, which is dependent only on the summation and distribution of weights, to a much richer dynamical space with many more degrees of freedom. Note that in the construction we describe here the notion of spatial summation in this dynamic perceptron is not being modeled since there is no physical geometry (morphology) to the nodes themselves. However, it is a further possible extension that would add to the dynamical repertoire of the model. 

Note also that this construction differs from the probabilistic extension of the deterministic version of the framework where one individual $v_i$ nodes are capable of activating $v_j$. In that formulation we assigned a probability distribution to the likelihood of activation of $v_j$ given a winning vertex $H_j[v^*_i] \leadsto v_j$. Here each $v_i \in H_j(v_i)$ contributes a component to the 'running' summation $\Sigma_r$.

We can compute when $\Sigma_r > \Sigma_T$ with a decaying memory by considering the refraction ratios $\Lambda_{ij}$ in the following way. We again make use of the well ordered set $\Lambda^o_{ij}$ from smallest to largest elements of $\Lambda_{ij} \in \Lambda^o_{ij}$(\emph{c.f.} equation \ref{eq:BigLSet}). 

Let $m$ index the order of the elements in the set $\Lambda^o_{ij}$. Then there exists a subset $\Lambda_M \subset \Lambda^o_{ij}$ such that $\Sigma_r \geq \Sigma_T$ when 
\begin{equation}
\Sigma_r = \sum_{m=1}^M w_{ij} - w_{ij} \cdot D_i(\bar{\tau}_{Mj} - \bar{\tau}_{ij}) 
\end{equation}
The function $D(t)$ is a positive asymptotic decay function evaluated at $(\bar{\tau}_{Mj} - \bar{\tau})$ with an asymptote $$\lim_{t \rightarrow C} D_i(t) = 1$$ where $C = (\bar{\tau}_{ij} - \bar{R}_j) + c$ for some positive constant $c$.

\begin{proof}
At the observation time $T_o$, we need to ask not (compute) which $v_i$ will win and activate $v_j$, $H_j[v^*_i] \leadsto v_j$, given the temporal evolutions of $\bar{\tau}_{ij} \in H_j$ and value of $\bar{R}_j$, but rather given the set of known weights $\{W_j$\}, temporal evolutions of $\bar{\tau}_{ij} \in H_j$, and value of $\bar{R}_j$ at $T_o$ compute which $v_i \in H_j$, i.e. which $v_i$, will result in $\Sigma_r \geq \Sigma_T$. 

Consider the set $\Lambda^o_{ij}$ with the order of elements indexed by $m$ such that $\Lambda_M \subset \Lambda^o_{ij} = \{(\Lambda_{ij})_m \in \Lambda^o_{ij}:m = 1, 2 \ldots M\}$ for $$\Sigma_r = \sum_{m=1}^M w_{ij} \cdot (v_i \in H_j) \geq \Sigma_T$$
This is the subset of $v_i$ connected to $v_j$ who's sum of weights equal or exceed the activation threshold of $v_j$. It is the subset of vertices that participate in activating $v_j$ assuming the summating contributing weights are persistently additive, in other words, do not decay. 

Of course, we need to also consider a decay of each contributing weight in time by a decay function $D_i(t)$ following the arrival of each event at $(\bar{\tau}_{ij} - \bar{R}_j)$ such that $w_{ij} < w_{ij,max}$ for times $t > (\bar{\tau}_{ij} - \bar{R}_j)$. $D_i(t)$ is a function of time versus a rate of decay normalized to unity. At the origin on this coordinate system $t = 0 = T_o$ is the time at which the network is observed or measured, which from a dynamical perspective effectively represents the value of $\bar{R}_{ij}$, the amount of time remaining in the relative refractory period of $v_j$. Then $D(\bar{\tau}_{ij} - \bar{R}_j)$ represents the number of time steps after $\bar{R}_{j}$ ends it takes for the signal from $v_i$ to arrive at $v_j$.

In the general case for each $v_i$ $D_i(t)$ must take the form $D(\bar{\tau}_{ij} - \bar{R}_j) = 0$ so that $w_{ij} = w_{ij,max}$ at $T_o$. The amount of decay for each $w_{ij}$, the time at which $D_i(t)$ is evaluated for each $v_i$, will depend on when the signal from $v_i$ arrives at $v_j$ relative to when the last contributing weight $w_{Mj}$ for vertex $v_{Mj}$ arrives at $(\bar{\tau}_{Mj} - \bar{R}_j)$ such that $\Sigma_r \geq \Sigma_T$. This effectively is a shifting of the decay function $D_i(t)$ dependent on when the signal from $v_i$ arrives before the last contributing weight sufficient to reach threshold arrives. Thus, $D_i(t)$ must therefore be evaluated at $t = [(\bar{\tau}_{ij} - \bar{R}_j) - (\bar{\tau}_{Mj} - \bar{R}_j)] = \bar{\tau}_{Mj} - \bar{\tau}_{ij}$.

Finally, given that $D(t)$ is defined as $$\lim_{t \rightarrow C} D_i(t) = 1$$ where $C = (\bar{\tau}_{ij} - \bar{R}_j) + c$ for some positive constant $c$, the decay of each summating weight scaled to the value of $w_{ij}$ must be $w_{ij} \cdot D_i(\bar{\tau}_{Mj} - \bar{\tau}_{ij})$. This ensures that $$D[(\bar{\tau}_{ij} - \bar{R}_j) + c] = w_{ij}$$ so that at $ t = (\bar{\tau}_{ij} - \bar{R}_j) + c$ the contributing value of weight from $v_i$ is zero. For each $v_i$ then its contribution to the summation $\Sigma_r$ minus the amount of decay from $w_{ij,max}$ at the arrival time $(\bar{\tau}_{ij} - \bar{R}_j)$ will be given by $$\Sigma_r = \sum_{m=1}^M w_{ij} - w_{ij} \cdot D_i(\bar{\tau}_{Mj} - \bar{\tau}_{ij})$$.
\end{proof}

It is important to note that we intentionally define only a generalized $D_i(t)$. The actual rate of change and kinetics of the decay function will of course have a profound impact on the dynamic spatial temporal summation of contributing weights to the perceptron, but the exact form of $D_i(t)$ could vary and indeed be optimized to a constructed network or specific task or objective. As such we leave it to the individual investigator to define. We also do not specify the form of the activation function of $v_j$, which can take on any form the investigator choses in order for $v_j$ to arrive at a decision as to whether to produce an output event in response to threshold being reached.  

Beyond the scope of this paper, on going research in our lab is exploring the use of geometric dynamic perceptrons in the construction of a new class of competitive refractory geometric dynamic artificial (recurrent) neural networks (gdANN's). One of the advantages gdANN's have over classical models is the ability to encode much more information for the same sized network due to a greater number of states as a result of the increase in the dimensionality endowed by the combinatorial dynamics of integrating signals.

\subsection*{Optimized information flows in geometric spatial temporal networks } \label{sec:optflow}
In this section we define a notion of efficient signaling between node pairs in a network within the context of our framework that naturally results from our theoretical arguments. In the next section we show how these results can be used to arrive at a unique interpretation for the prevalence of small world networks. Given an effective refractory period $\bar{R}_j$ and effective delay time or latency $\bar{\tau}_{ij}$ along an  edge $e_{ij}$, the condition for the winning vertex $v_i \in H_j(v_i)$ that achieves activation of $v_j$, i.e. $H_j[v^*_i] \leadsto v_j$, is dependent on the $\lim_{\bar{\tau}_{ij} \rightarrow \bar{R}^+_j} \Lambda_{ij} = \bar{R}_j/\bar{\tau_{ij}}$.
This implies a balance between how fast information is capable of propagating through the network relative to how quickly its nodes can process incoming signals. When a mismatch between network geometry and dynamics exists, it can render the nodes unable to process any information at all or can result in highly inefficient network dynamics. We have previously shown that if  signaling speeds $s_{ij}$ are too fast, or equivalently, if the latencies $\bar{\tau}_{ij}$ are too short compared to the amount of time a node requires to process an input, generate an output, and recover to a state in which it can respond again, the network will not be able to sustain internal recurrent activity \cite{buibas}.  If $s_{ij}$ is too slow or the set of $\bar{\tau}_{ij}$ too long then the network will be inefficient in the sense that it has the potential for faster dynamic signaling that is not being realized. Time, as a resource, is being wasted in such a network. In the following sections we formalize these intuitive concepts.

\subsubsection*{Optimized bounded refraction ratio} 
We derive upper and lower bounds on the signaling dynamics, which we in turn use to define what we mean by an optimized refraction ratio between connected vertices $v_i$ and $v_j$. \newline
 
\textbf{Theorem 4. Optimized refraction ratio theorem.}\label{theo:opt}
Let $G=(\bar{V},\bar{E})$ represent a complete geometric graph model of a network consisting of subgraphs $H_j(v_i)$ such that all $v_{i} \in H_j(v_i)$, contain directed edges $H_j(\bar{E})$ into vertex $v_j$. For each $v_i v_j$ vertex pair with a signaling speed $s_{ij}$ between $v_i$ and $v_j$, the optimal refraction ratio $[ \Lambda_{ij} ]_{opt}$ is bounded by
\begin{subequations} \label{eq:opt}
\begin{align}
& [ \Lambda_{ij} ]_{opt} = \lim_{\tau_{ij} \rightarrow R_{j}^+} \Lambda_{ij} \text{ when } \phi_j \text{ and } \delta_{ij} = 0 &&\text{ [Upper bound ]}  
\\ & [ \Lambda_{ij} ]_{opt} \Rightarrow \lim_{\delta_{ij} \rightarrow -\phi_j^+} \Lambda_{ij} \text{ when } \phi_j = R_j &&\text{ [Lower bound]}
\end{align}
where $\tau_{ij}$ is the absolute signaling delay on the edge $e_{ij}$ and $R_j$ is the absolute refractory period for $v_j$.
Given these bound then, an optimized refraction ratio will be such that 
\begin{equation}
[ \Lambda_{ij} ]_{opt} = \frac{\bar{R}_j}{\bar{\tau}_{ij}} \rightarrow 1
\end{equation}
\end{subequations}

\begin{proof}
 The necessary condition for the activation of $v_j$ by $v_{i} \in H_j$ is $\bar{\tau}_{ij} > \bar{R}_j$. By equation \ref{eq:rbar} $\bar{R}_j = R_j - \phi_j \text{ where } 0 \leq \phi_j \leq R_j$, which implies that $0 \leq \bar{R}_j \leq R_j$. $\bar{R}_j$ is bounded by its very construction. The absolute lower bound on $\bar{R}_j$ implies that activation of $v_j$ by a $v_{i} \in H_j$ will be achieved when $\bar{\tau}_{ij} > 0$, and the absolute upper bound implies that $\bar{\tau}_{ij} > R_j$. But note how $\bar{\tau}_{ij}$ can always achieve these bounds independent of $\tau_{ij}$ for a given $v_i$ $v_j$ pair at some observation time $T_o$ because by equation \ref{eq:taubar} $\delta_{ij} \in \mathbb{R}$, i.e. any vertex $v_i$ can activate $v_j$ independent of the absolute latency $\tau_{ij}$ by delaying the initiation of an output signal at $v_i$ long enough if $\tau_{ij}$ is too short or initiating a signal at $v_i$ prior to $T_o$ if $\tau_{ij}$ is too long. However, by theorems 2 and 3, $\bar{\tau}_{ij}$ need only be slightly larger than $\bar{R}_j$ in order to successfully signal $v_j$: $\bar{\tau}_{ij} \rightarrow \bar{R}_j^+$. Because $\bar{R}_j$ is naturally bounded by $0 \leq \bar{R}_j \leq R_j$, it follows intuitively that the optimal signaling condition will be given by $\tau_{ij} \rightarrow \bar{R}_j^+$ for values of $\delta_{ij}$ not too smaller than zero or not too greater than zero in order to meet the condition that $\tau_{ij} \rightarrow \bar{R}_j^+$ while avoiding compensation by $\delta_{ij}$. In other words, the response dynamic range for any $v_j$ will always be bounded by the limits of $\bar{R}_j$ in the sense that these limits determine the temporal properties of when $v_j$ can actively participate in network signaling and when it cannot. Ultimately, of course, this is a function of $v_j$'s internal dynamics, which in turn determines $R_j$ and $\bar{R}_j$. No value of $\tau_{ij}$ need be much greater than $R_j$ for any $v_i$ with a directed edge $e_{ij}$ into $v_j$. When the condition $\tau_{ij} \rightarrow \bar{R}_j^+$ is met, it ensures that such a $\tau_{ij}$ is guaranteed to be able to operate over the entire response dynamic range of $v_j$, i.e. all values of $\bar{R}_j$ (see also the corollary results below).

For the upper bound this optimized boundary condition will occur when $\tau_{ij} \rightarrow R_j^+$ when $\phi_j$ and $\delta_{ij} = 0$ because it represents the upper achievable limit for $\bar{R}_j$ (when $\phi_j = 0$) and forces the optimal condition that $\tau_{ij} \rightarrow R_j^+$ without compensating with $\delta_{ij}$. For the lower bound the optimal condition is given by $\bar{\tau}_{ij} \rightarrow 0^+ \Rightarrow \delta_{ij} \rightarrow -\phi_j^+$ when $\phi_j = R_j$, since when $\phi_j = R_j \Rightarrow \bar{R}_{ij} = 0$. Forcing the condition that $\phi_j = R_j$ implies that $\tau_{ij}$ on its own is capable of meeting the lower bound without compensation by $\delta_{ij}$. Formally, we can define an optimized bound as $|\tau_{ij} - R_j| < \epsilon$ for some bounded error $\epsilon$.  If $\bar{\tau}_{ij}$ is too short, either because the path length of $e_{ij}$ is too short or $s_{ij}$ is too fast, this implies that given $R_j$, $\bar{\tau}_{ij} \nrightarrow R_j^+$ if $\delta_{ij} = 0$. To achieve the lower bound it would require $\delta_{ij} < 0$ so that $\bar{\tau}_{ij} < \tau_{ij}$. To achieve the upper bound it would require $\delta_{ij} > 0$ so that $\bar{\tau}_{ij} > \tau_{ij}$. If $\bar{\tau}_{ij}$ is too long, either because the path length of $e_{ij}$ is too long or $s_{ij}$ is too slow, this implies that given $R_j$, $\bar{\tau}_{ij} \nrightarrow 0^+$ if $\delta_{ij} \rightarrow -\phi_j^+$. When these constraints are met, it ensures that $\bar{R}_j/\bar{\tau}_{ij} \rightarrow 1$.
\end{proof}

For the lower bound the important condition is that $\bar{\tau}_{ij} \rightarrow 0^+$ when $\delta_{ij} \rightarrow -\phi_j^+$. It is trivial what $\delta_{ij}$ is, since this condition will always be met when $\delta_{ij} = -\tau_{ij}$. But for the upper bound the important condition is that $\tau_{ij} \rightarrow R_j^+$ when $\delta_{ij} = 0$, which implies that in every case $\bar{\tau}_{ij} = \tau_{ij}$. 

What these bounds imply is that if $v_i$ satisfies the bounding conditions, it will always be within a range where it could 'win' and activate $v_j$ for any value of $\bar{R}_j$ at any observation time $T_o$ and time of signaling initiation of $v_i$ $t_i$. The given $v_i$ may not of course always 'win' in activating $v_j$ but it is insured to be as efficient as possible, as efficient as any other node in its signaling of $v_j$ over all values of $\bar{R}_j$. If $R_j$ is the same for all nodes in the network, $R_j = R$ $\forall j \in G(\bar{V},\bar{E})$, this in effect bounds the dynamic window over which all network dynamics, that is, all temporal information (signaling) processes, should occur on that network. Explicitly, this dynamic window is given by $\bar{R}_j = R_j - \phi_j \text{ for } 0 \leq \phi_j \leq R_j$. So there is no need or reason for any $\tau_{ij}$ to go far beyond this dynamic window in order to satisfy the optimality condition $\bar{\tau}_{ij} \rightarrow \bar{R}_j^+$. This is in essence the intuitive basis of using the bounds derived in theorem \ref{theo:opt} to define optimized signaling or information flow between node pairs. 

Note that we must keep the explicit condition that $\phi_j = R_j$ because that forces $\Lambda_{ij} = 0$ only when $\delta_{ij} \rightarrow -\phi^+$. Otherwise, in the general case any value of $\bar{\tau}_{ij} >> 0$ will result in $\Lambda_{ij} = 0$ for any value of $\bar{R}_{j}$.

We can also make the following statement. Given the conditions for lower and upper bounds in theorem \ref{theo:opt}, let $[ \delta_{ij} ]_{upper}$ denote the value that $\delta_{ij}$ must take in order to achieve the optimal upper bound condition for some value of $\tau_{ij}$ that does not necessarily satisfy $[ \Lambda_{ij} ]_{opt}$. Similarly, let let $[ \delta_{ij} ]_{lower}$ denote the value that $\delta_{ij}$ must take in order to achieve the optimal lower bound condition. In every case, the relationship between $[ \delta_{ij} ]_{upper}$ and $[ \delta_{ij} ]_{lower}$ is given by
\begin{equation*}
[ \delta_{ij} ]_{lower} = [ \delta_{ij} ]_{upper} - R_j \\
\end{equation*}

\begin{proof}
The condition for the upper bound is $\tau_{ij} \rightarrow R_j^+$ for $\delta_{ij} = 0$ (and $\phi_j =0$). With a bit of abuse of the relational terms, since $\bar{\tau}_{ij} = \tau_{ij} + \delta_{ij}$, under these conditions if $\tau_{ij} \neq R_j$ then $\bar{\tau}_{ij} \nrightarrow R_j^+$. For a given $\tau_{ij}$ and for a known or measurable $R_j$, $[ \delta_{ij} ]_{upper} = -(\tau_{ij} - R_j)$, since this is what the value of $\delta_{ij}$ would have to be in order to achieve the optimal condition. For the lower bound $\bar{\tau}_{ij} \rightarrow 0^+$ when $\delta_{ij} = -\phi_j$ given that $\phi_j = R_j$. Thus, if $\tau_{ij} -\phi_j \equiv \tau_{ij} -R_j > 0$, or more correctly if $\bar{\tau}_{ij} \nrightarrow 0^+$, it implies that $[ \delta_{ij} ]_{lower} = -\tau_{ij}$ would be needed to meet the lower bound optimality condition. The difference between $[ \delta_{ij} ]_{lower}$ and $[ \delta_{ij} ]_{upper}$ is therefore $- \tau_{ij} - [- (\tau_{ij} - R_j)] = - R_j$. Thus, $[ \delta_{ij} ]_{lower} = [ \delta_{ij} ]_{upper} - R_j$. 
\end{proof}

If a signal $v_i \in H_j(v_i)$ characterized by a latency $\tau_{ij}$ on the edge $e_{ij}$ is able to achieve either the optimal upper bound or optimal lower bound as defined in theorem 4, then it is guaranteed to be able to achieve the other optimal bound.

\begin{proof}
Asking if a signal capable of achieving the upper bound can also achieve the lower bound is equivalent to asking if $\bar{\tau}_{ij} \rightarrow 0^+$ when $\tau_{ij} = R_j$. But the condition for the lower bound is $\bar{\tau}_{ij} \rightarrow 0^+$ when $\delta_{ij} = -\phi_j$. Substituting for these explicit variables we arrive at 
\begin{align*}
\bar{\tau}_{ij} &= \tau_{ij} + \delta_{ij}\\
&= R_j - \phi_j
\end{align*}
but since $\phi_j = R_j$ for the lower bound, it implies that $\bar{\tau}_{ij}  = 0$, or more appropriately, $\bar{\tau}_{ij}  \rightarrow 0^+$. 

Asking if a signal that satisfies the optimality condition for the lower bound can also achieve the upper bound is equivalent to asking if $\bar{\tau}_{ij}  \rightarrow R_j^+$ when $\delta_{ij} \rightarrow -\phi_j$. Similarly, 
\begin{align*}
\bar{\tau}_{ij} &= \tau_{ij} + \delta_{ij}\\
0 &= \tau_{ij} - R_j\\
\tau_{ij} &= R_j
\end{align*}
which implies that the optimality condition for upper bound is satisfied. 
\end{proof}

The definition of an optimally efficient network then follows: In every case, as a function of the effective refractory period $\bar{R}_j$ and effective delay time $\bar{\tau}_{ij}$ along the edge $e_{ij}$, the condition for the winning vertex $v_i$ that achieves activation of $v_j$, i.e. $H_j[v^*_i] \leadsto v_j$, is dependent on the $\lim_{\bar{\tau}_{ij} \rightarrow \bar{R}^+_j} \forall v_i \in G(\bar{V},\bar{E})$. When this condition is satisfied for all edges $e_{ij} \in \bar{E}$, i.e. $\bar{E}=\{e_{ij}\}$, for all $v_i$$v_j$ node pairs by the upper and lower bound definitions for $[ \Lambda_{ij} ]_{opt}$ (equation \ref{eq:opt}) such that $\Lambda_{ij} =  \bar{R}_j / \bar{\tau}_{ij} \rightarrow 1$, the network is optimally efficient, i.e. $[ \Lambda_{ij} ]_{opt} \forall v_i \in G(V,E)$. This is equivalent to requiring the condition $|\tau_{ij} - R_j| < \epsilon$ $\forall v_i \in G(\bar{V},\bar{E})$ for some arbitrarily small value of $\epsilon$.

A consequence of this is that for a network to meet this strict definition it must exhibit a lattice structure. Given constant values of $|t_{j}|$ and $R_j$ $\forall \bar{V}$ and $s_{ij}$ $\forall e_{ij} \in \bar{E}$ in $G(\bar{V},\bar{E})$, optimized signaling efficiency at the network scale is only achievable when $d_{ij} = |e_{ij}| = C$ $\forall \bar{E} \in G(\bar{V},\bar{E})$, where $C$ is a constant. This geometrically implies that the network must exhibit a lattice structure. 

\begin{proof}
Given the assumptions that $|t_{j}|$ and $R_j$ $\forall \bar{V}$ and $s_{ij}$ $\forall e_{ij} \in \bar{E}$ in $G(\bar{V},\bar{E})$, assume optimal signaling between any arbitrarily chosen $v_Iv_J$ pair such that $\Lambda_{IJ} \rightarrow 1^+$. This then implies that
\begin{equation*}
\Lambda_{ij} = \frac{R_J - \theta_J}{\tau_{IJ} - \delta_{IJ}} = \frac{R_J - \theta_J}{\frac{d_{IJ}}{s_{IJ}} + \delta_{IJ}} \rightarrow 1^+
\end{equation*}
For the lower bound $\phi_i$ and $\delta_{ij} = 0$, so the condition for optimal signaling will be $R_j \cdot (s_{ij}/d_{ij}) \rightarrow 1^+$. This means that for any $v_iv_j$ pair such that $d_{ij} \neq d_{IJ}$ $\Lambda_{ij} \nrightarrow 1^+$.

For the upper bound $\phi_j =\delta_{ij}$, so 
\begin{equation*}
\Lambda_{ij} =  \frac{R_j - \delta_{ij}}{\frac{d_{ij}}{s_{ij}} + \delta_{ij}} \rightarrow 1^+
\end{equation*}
By corollary \ref{cor:opt1}, let $[ \delta_{ij} ]_{lower} = [ \delta_{ij} ]_{upper} - R_j := K$. Then
\begin{equation*}
\Lambda_{ij} =  \frac{R_j - K}{\frac{d_{ij}}{s_{ij}} + K} \rightarrow 1^+
\end{equation*}
Then evaluating again for the lower bound given the substitution yields the same result as above, such that $\Lambda_{ij} \nrightarrow 1^+$.
\end{proof}

A subtle but important consideration is that while this definition of optimal network efficiency provides a strict criteria of optimized network signaling by evaluating the optimality of signaling across all node pairs independently, it is not necessarily equivalent to optimized network dynamics. In other words, a pure lattice structure may not be the best geometric connectivity to optimize the dynamical flow of information across a network in support of some function or behavior or learning the network is intended to do. This definition is strictly mathematical.

\subsection*{The pervalence of the small world network topology explained by the refraction ratio}
In 1998 Watts and Strogatz published their seminal paper introducing small world networks, and suggested that this topology could be pervasive across both natural and engineered networks \cite{Watts:1998vc}. The small world network connectivity structure lives between a completely random network and a regular lattice. It provides an opportunity for nodes that would normally not be connected to be connected, resulting in a 'short circuiting' of dynamical behaviors and communication between different parts of the network that would normally not be in such immediate and direct contact. A key observation was that the transition to a small world topology from a regular lattice is essentially undetectable at the local scale, but can have significant effects on the dynamics and the spread of information. 

What is less obvious though, is \emph{why} these networks are so prevalent. In other words, why it is that this connectivity topology re-emerges across many natural and engineered systems is not obvious. But there must be some deep underlying process beyond the physics of the individual systems. Our theory suggests one possible explanation. One possible interesting interpretation of why such few long range connections have such a significant effect on network dynamics is the following: assuming consistent internal node dynamics and a constant signaling speed on all edges, sparse long range random re-wiring events sufficient to produce a meaningful effect on network dynamics, as shown in \cite{Watts:1998vc}, can be constructed from an optimally efficient regular lattice network with essentially no effect on a loss of optimization or efficiency as defined by Theorem 4 above.

To see this, consider first the clustering coefficient in a small world network, $C(p)$, which is a measure of the degree of local connectivity. It reflects the deviation from a regular lattice network at the local scale. (See Fig. 2 in  \cite{Watts:1998vc} for a formal definition.) $C(p)$ is nearly constant and unchanging from a lattice network until about $p \approx 0.01$; $C(p)/C(0) \approx 1$, where $C(0)$ represents a value of $p=0$, i.e. no random rewiring and a lattice network structure. In contrast, the characteristic path length $L(p)$ for $p=0.01$ as a ratio of the characteristic path length of a lattice network $L(0)$ is $L(p)/L(0) \approx 0.2$, indicating a significant degree of long range re-wiring events at the scale of the whole network. $L(p)$ is a measure of the shortest path between two vertices, averaged over the entire network. Consider a network optimization cost function $C_N$ associated with a deviation of $\Lambda_{ij}$ for a $v_i v_j$ node pair. Computing $C_N$ for all node pairs results in the cost of deviation from a lattice network for the entire network being considered (see Methods below). If we compare how $C_N$ changes as a function of $p$ we observe that at the critical transition probability $p = 0.01$ there is essentially a negligible change in $C_N$ relative to an optimized regular lattice network (Fig. 1). We considered $C_N$ for families of small world networks with increasing random re-wirings starting from a regular lattice network that had a ratio $\Lambda_{ij} = 1.2$ for all vertex pairs. We then investigated how $C_N$ changed as a function of increasing $p$ for small world networks where the random re-wirings produced signaling delays that resulted in 2x, 10x, and 20x $>  [ \Lambda_{ij} ]_{opt}$ for re-wired vertex pairs. In all cases, $\Delta C_N < 0.019$ at $p = 0.01$ even though $\Lambda_{ij} =$ 1.667 (for the curve labeled 2x; see inset in Fig. 1), 8.333 (curve labeled 10x), and 16.667 (curve labeled 20x) for the fraction of re-wired edges. In other words, even though the signaling efficiency for each of the three networks (by design) progressively deviated from $[ \Lambda_{ij} ]_{opt}$, the additional deviation from $[ \Lambda_{ij} ]_{opt}$ introduced by random re-wirings associated with a small world network, measured by the cost function $C_N$ was essentially negligible at re-wiring probabilities (i.e. $p=0.01$) that produce significant effects on network dynamics attributed to the small world topology. We then explored how $C_N$ changes at fixed $p$ values as a function of increasing re-wiring signaling delays, expressed again as $x$ times greater than the signaling speed associated with $[ \Lambda_{ij} ]_{opt}$ (Fig. 2). The change in $C_N$ was linear but with different slopes, reflecting the value of $p$. While there is an increase in $C_N$ associated with re-wired edges that have signaling speeds that progressively move $\Lambda_{ij}$ away from $[ \Lambda_{ij} ]_{opt}$, as would be expected, the change is linear and rather flat. Lastly, we explored how $C_N$ changed as a function of $p$ for different deviations from $[ \Lambda_{ij} ]_{opt}$ for the starting regular lattice network (Fig. 3). Increasing the initial deviation from $[ \Lambda_{ij} ]_{opt}$ for all vertex pairs in the starting lattice network did not change the dynamics of how $C_N$ varied, but did affect the starting value of $C_N$ across all values of $p$, as would be expected. 

Small world networks represent a connectivity class which can be designed to display arbitrarily optimal signaling dynamics with essentially negligible deviation from regular lattice networks while simultaneously displaying sufficient long range random edge re-wirings (at $p = 0.01$) to produce a significant impact on the dynamics of the network. It makes sense for both natural and engineered networks to take advantage of this topological structure, since it provides a simple set of construction rules to produce tailored effects on dynamics and the propagation of information through the network (via the re-wiring probability $p$) with essentially no resource costs in doing so. The analysis we describe here suggests a unique interpretation of why and how small world networks are able to achieve such dramatic effects on network dynamics. 

\begin{figure}
\begin{center}
\includegraphics[width=3.5in]{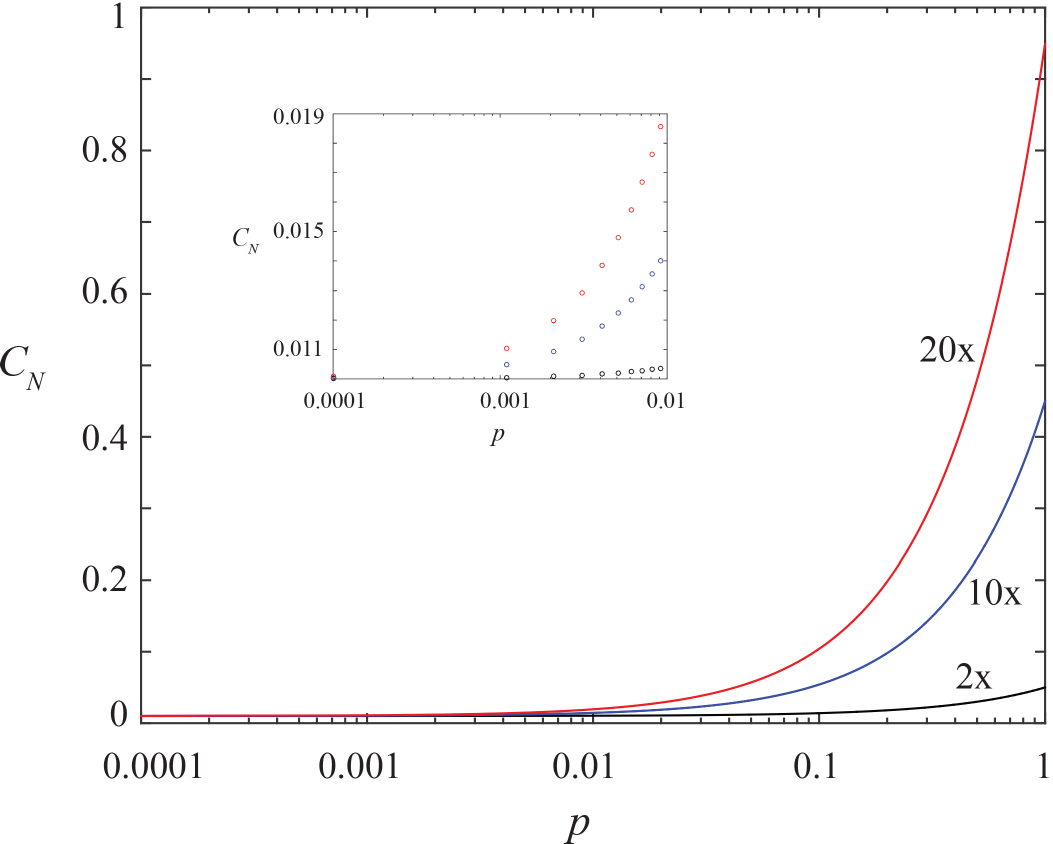}
\caption{\textbf{Network cost function $C_N$ as a function of small world re-wiring probabiltiy $p$ for three networks with differing deviations of dynamic signaling optimality for re-wired edges.} The starting lattice network was considered optimally efficient with $\Lambda_{ij} = 1.2 :=  [ \Lambda_{ij} ]_{opt}$. Randomly re-wired long range connections had signaling delays that resulted in 2x, 10x, and 20x $> [ \Lambda_{ij} ]_{opt}$. Inset: Magnified $C_N$ scale up to $p= 0.01$.}\label{fig:cn1}
\end{center}
\end{figure}

\begin{figure}
\begin{center}
\includegraphics[width=3.5in]{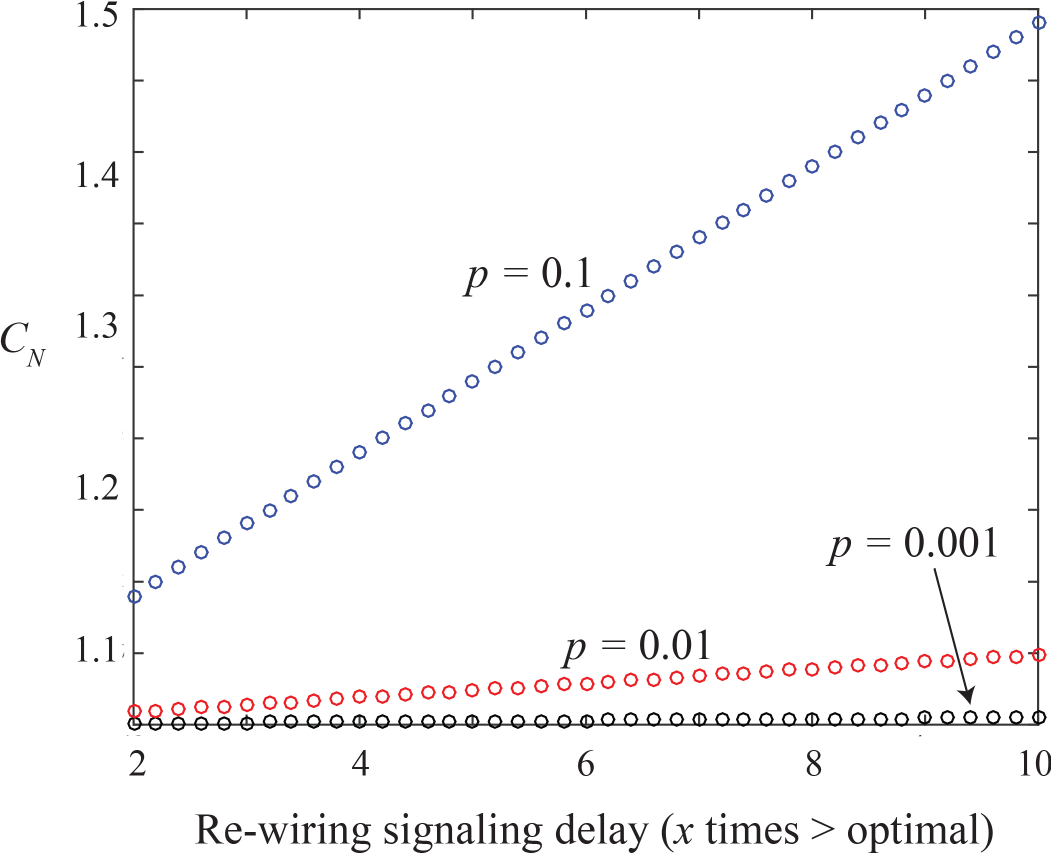}
\caption{\textbf{Network cost function $C_N$ as a function of increasing re-wiring deviation from $[ \Lambda_{ij} ]_{opt}$ for small world networks.} $C_N$ was computed for value of $p = 0.001$, 0.01, and 0.1.}\label{fig:cn2}
\end{center}
\end{figure}

\begin{figure}
\begin{center}
\includegraphics[width=3.5in]{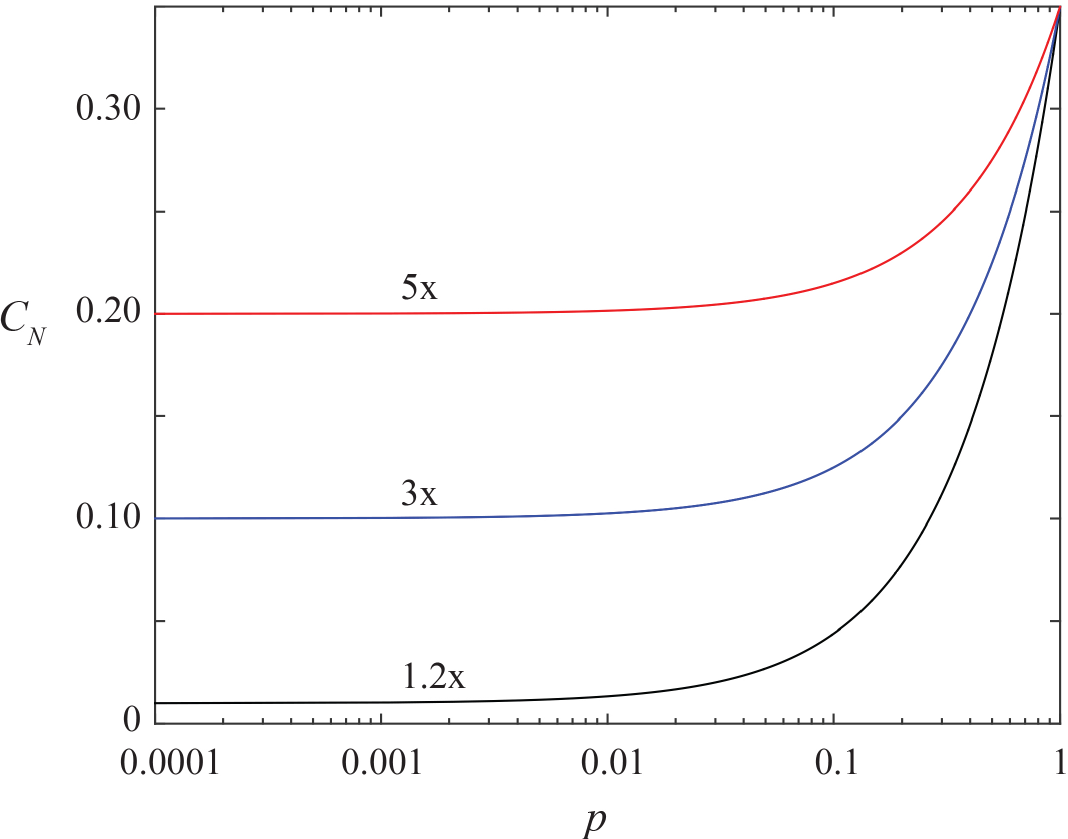}
\caption{\textbf{Network cost function $C_N$ as a function of small world re-wiring probabiltiy $p$ for three networks with differing deviations from $[ \Lambda_{ij} ]_{opt}$ for the starting lattice network.} Signaling delays for the starting lattice network were 1.2, 3x and 5x $>$ than $[ \Lambda_{ij} ]_{opt}$.}\label{fig:cn3}
\end{center}
\end{figure}

\section*{Methods}
We considered the small world topology by first defining a cost function associated with deviation of $\Lambda_{ij}$ for a $v_i v_j$ node pair. We can define a simple cost function as $C_{ij} = |\tau_{ij}  - R_j|$ for every connected vertex pair and then average over the entire network: $C_{N} = \frac{\sum^K_{k = 1} C_k}{K} \text{ for } k = 1, 2, 3 \dots , K \text{ edges }$ for each connected $v_i v_j$ pair. A regular lattice network with probability $p$ of long range random re-wirings equal to zero (see  \cite{Watts:1998vc} for details), can be made to exhibit arbitrarily near optimal signaling dynamics approaching $[ \Lambda_{ij} ]_{opt}$ since the speed of signal propagation can balance the length of the edges such that they match any given internal node dynamics. Because it is a regular lattice network, with all edges being equal, this will apply to the entire network and $C_N \rightarrow 0$. 

\section*{Discussion}
In this paper we present an intuitively simple framework that describes the competing dynamics of signaling and information flows in geometric networks (as we define them above) derived from foundational principles of biological cellular neural signaling. These results suggest an explanation for how the interplay between strictly local geometric and temporal process at the scale of individual interacting nodes gives rise to the global behavior of the network. It is important to emphasize that the response of each vertex is causally independent from whatever all the other vertices in the network are doing. Computing $y_j$ at any instantaneous time is only dependent on the internal dynamics of $v_j$, which in turn determines $R_j$, and when signals arrive from (competing) input vertices into $v_j$. This allows the independent computation of the interacting states of any $v_iv_j$ vertex pair at the observation time $T_o$. The computation of $y_j$ is not dependent on any 'average' metric of the state or behavior of the network as a whole, or on any statistical probability densities associated with the frequency of occurrence of events such as in Markovian processes. Computing in parallel all $v_i v_j$ state pairs reports back the state of the overall network. 

Some theoretical aspects of the framework were intentionally left open in order to provide sufficient application specific flexibility in how the theory can be used. For example, the situation where there is a tie with regards to two signals reaching a node $v_j$ at exactly the same moment. We did not attempt to explicitly account for this because the details of how such a tie is broken is a property of the system itself. Consider as an example Watt's and Strogatz's simulation of the spread of infections on small world networks (see their Fig. 3 in \cite{Watts:1998vc} and its accompanying text). The tie breaking rule in that case is that if two vertices are at the same time step attempting to infect the same $v_j$ there is no competition at all and no tie breaking rule is needed. Either $v_j$ will be infected given the respective probabilities of each $v_i$ (or the same probability for a given $r_i$ equal for the whole network) or it will not. The competitive refractory dynamics in this model only exists at the level of nodes attempting to reach uninfected nodes at a given time step first, and do not directly compete with themselves when they infect at the same time.

There is a growing literature on the topic of spatio-temporal networks \cite{George:2013b,Holme:2013p}.  Temporal networks are networks modeled as temporally dynamic graphs that change with time. Edges have temporal properties with time dependent attributes such as signaling latencies that determine the signaling or information flow dynamics on the network. Spatio-temporal networks extend this notion to include vertices that have physical geometric representations in space (or the plane). In our work we also considered the geometry of the edges as a third explicit component that contributes to the dynamics. In this context, temporal activation times on the edges and the spatial positions of vertices are an extension of the primitive notion of the connectivity of a graph. The underlying structural connectivity or topology of the network represents the totality of all the possible pathways over which signals or information can flow within the network, the total solution space of the graph. But at any instantaneous moment in time, such as a specific instance at which the network is observed or measured, only a subset of these pathways will be active, dependent on the temporal dynamics of the set of signals on the edges and their ability to activate the vertices they are connected to. 

Temporal networks and spatio-temporal networks are of relevance to many natural, engineering, and technological systems, because the conceptual model they provide naturally maps to many real world physical systems. Examples include transportation systems, communication networks, social networks, the spread of infections diseases (including computer viruses and malware), physical-chemical systems such as the interactions of particles in solution and diffusion, and -omics type biological networks of molecular and genetic interactions. Of particular interest to us is the signaling dynamics responsible for computation in biological neural networks across scales of organization (e.g. dendritic trees, neural circuits, or the interaction between different brain regions) constrained by the underlying structural connectome appropriate for each scale. Yet, despite their increasingly recognized relevance and importance, and while intuitively appealing from a modeling perspective, existing theories of temporal networks and spatio-temporal networks are still very much in their infancy. Most of the existing literature has focused on descriptive theory, definitions of concepts and notation, and various functional metrics. There are still very few applications, concrete results and uses, predictions, or theoretical predictive analyses. There is an almost exploratory quality to the existing literature. In contrast, we were primarily interested in concrete theoretical arguments, which we have previously argued is critical to advancing our understanding of the brain beyond descriptive models \cite{Silva:2011gs}.

\section*{Acknowledgements }
I am very grateful to Prof. Fang Chung (Department of Mathematics, UC San Diego), Prof. Shankar Subramaniam (Department of Bioengineering, UC San Diego), Dr. Marius Buibas, and (soon to be) Dr. Vivek George for their input and many hours of fruitful discussions. This work was supported by grants 63795EGII and N00014-15-1-2779 from the Army Research Office (ARO), United States Department of Defense.

\section*{Author contributions statement}
All parts of this work and manuscript were done by G.S.

\bibliographystyle{bmc-mathphys}

\end{document}